\documentclass[a4paper]{article} 

\author{Jacques Smulevici \footnote{University of Cambridge,
Department of Applied Mathematics
and Theoretical Physics, Wilberforce Road, Cambridge CB3 0WA United Kingdom}}

\title{Strong cosmic censorship for $T^2$-symmetric spacetimes with positive cosmological constant and matter}

\usepackage{amsmath}
\usepackage{graphicx,color}
\usepackage{latexsym}
\usepackage{amsmath,amsthm}
\usepackage{amssymb}

\newtheorem{proposition}{Proposition}
\newtheorem{theorem}{Theorem}
\newtheorem{lemma}{Lemma}
\newtheorem{corollary}{Corollary}

\newcommand{\demi}{\frac{1}{2}}
\newcommand{\demir}{\frac{1}{2r}}
\newcommand{\demirr}{\frac{1}{2r^2}}
\newcommand{\quartrr}{\frac{1}{4r^2}}

\newcommand{\drond}[2]{\frac{\partial #1}{\partial #2}}
\newcommand{\quart}{\frac{1}{4}}
\newcommand{\comment}[1]{}

\begin{document}

\maketitle

\begin{abstract}
We address the issue of strong cosmic censorship for $T^2$-symmetric spacetimes with positive cosmological constant. In the case of collisionless matter, we complete the proof of the $C^2$ formulation of the conjecture for this class of spacetimes. In the vacuum case, we prove that the conjecture holds for the special cases where the area element of the group orbits does not vanish on the past boundary of the maximal Cauchy development.
\end{abstract}

\tableofcontents

\section{Introduction}

Strong cosmic censorship is one of the most interesting and important conjectures of mathematical relativity. The $C^2$ formulation of this conjecture states that the maximal Cauchy development of generic compact or asymptotically flat initial data for suitable Einstein-matter systems is locally inextendible as a $C^2$ regular Lorentzian manifold \cite{ch:givp}.

Some results have been obtained when the initial data are required to have prescribed isometries. Among the symmetric spacetimes which have been studied are the $T^2$-symmetric solutions which constitute a class of spacetimes admitting a torus action, the $T^3$-Gowdy solutions which are special cases of $T^2$-symmetric solutions and the surface symmetric solutions which constitute a class of spacetimes admitting plane, hyperbolic or spherical symmetry.
In particular, the conjecture was proved for $T^3$-Gowdy solutions in the vacuum \cite{hr:scct3}, $T^2$-symmetric and surface symmetric spacetimes with collisionless matter and vanishing cosmological constant ($\Lambda=0$) \cite{dr:t2, dr:ss}.
For the $T^3$-Gowdy vacuum solutions, the proof of the conjecture relies on a detailed asymptotic analysis of solutions near the past boundary whereas for the $T^2$- and surface symmetric cases with collisionless matter, it relies on a rigidity property of possible horizons and on the characteristic properties of the Vlasov matter. If the proofs are very different in nature, an understanding of the behaviour of the area of the orbits of symmetry is always necessary.

For $T^2$-symmetric spacetimes with various matter models, it is known that the area of the orbits of symmetry will take all values in $(r_0, +\infty)$ for some $r_0 \ge 0$ \cite{bicm:gft2, arw:caft2v, ci:t2pcc}\footnote{More is actually known since the existence of global areal foliations where $r$ is considered as a time coordinate has been established. Note also that the existence of constant mean curvature foliations has been obtained for these spacetimes.}. In particular, the unboundedness of the area of the orbits of symmetry implies inextendibility in the expanding direction \cite{dr:iecs}. Therefore it is sufficient to study the contracting direction to complete the proof of strong cosmic censorship for these models. For the $T^2$-symmetric spacetimes (including the Gowdy case) with $\Lambda=0$ and with or without Vlasov matter, it was later shown that $r_0=0$  (except for some flat Kasner cases) \cite{iw:ast2, mw:ast2v}.

Less is known for symmetric spacetimes with positive cosmological constant ($\Lambda > 0$). In the surface symmetric case with collisionless matter, the conjecture was proved for the hyperbolic or plane symmetric cases only, some obstructions remaining to show strong cosmic censorship for the spherical symmetric case \cite{dr:ss}. One of the main difficulties comes from the lack of information about the behaviour of the area of the group orbits close to the past boundary since for the $T^2$-symmetric spacetimes with $\Lambda > 0$, it is not known whether generically $r_0=0$ and in the case of spherical symmetry with $\Lambda > 0$, the area of the orbits of symmetry is not constant on the past or the future boundary \cite{dr:ss, ci:t2pcc}. Moreover, in the surface symmetric case with spherical symmetry, another difficulty comes from the so-called extremal case ($r_0=\frac{1}{\sqrt{\Lambda}}$).

One of the goals of this article is to show that even without a complete understanding of the behaviour of the area of the orbits of symmetry in the contracting direction, we can still prove $C^2$ inextendibility for $T^2$-symmetric spacetimes with $\Lambda > 0$ and collisionless matter, we need only to proceed by dichotomy. Indeed, the proof of M. Dafermos and A. Rendall of the conjecture for $T^2$-symmetric spacetimes with collisionless matter and $\Lambda=0$ can also be applied for $\Lambda > 0$ under the extra assumption that the area of the group orbits vanishes on the past boundary, namely $r_0=0$. In order to complete the proof of strong cosmic censorship for this class of spacetimes,  it is therefore sufficient to restrict ourselves to the cases where $r_0>0$. This is the content of Theorem \ref{th:vla} which states that \emph{$T^2$-symmetric spacetimes with $\Lambda > 0$, non-vanishing collisionless matter and for which $r_0>0$  are $C^2$ inextendible.}

Understanding extendibility or inextendibility of vacuum spacetimes is generally thought to be harder as is well illustrated by the difficult analysis of the vacuum Gowdy spacetimes. Theorem \ref{maintheorem} states that \emph{vacuum $T^2$-symmetric spacetimes with $\Lambda >0$ and for which $r_0>0$ on the past boundary are generically $C^2$ inextendible.} Thus, Theorem \ref{maintheorem} reduces the issue of strong cosmic censorship in this class of spacetimes to the cases where the area of the symmetry orbits takes all positive values. This is an unexpected result as it was thought that proving $r_0=0$ was a necessary step towards a proof of generic inextendibility.

Assuming the spacetime to be extendible, we will first use the $C^1$ extension of the Killing fields to the Cauchy horizon \cite{dr:iecs} in order to infer initial data in a bounded null coordinate system for some energy functions defined for $T^2$-symmetric spacetimes. Analysing these energy functions along characteristics, we will obtain pointwise estimates in a fundamental domain of the universal cover of the spacetime. These pointwise estimates mean that the part of the Cauchy horizon which coincides with the past boundary of the fundamental domain is then everywhere regular. Using this regularity, we will evaluate the Raychaudhuri equation on the past boundary in order to obtain a certain rigidity of the horizon.
In the vacuum case, this rigidity will be translated in a new set of initial data on the past boundary for the same energy functions. With these new initial data, we will carry another analysis of the energy functions along characteristics in order to extend our bounds to the tip of the universal cover. As before, the regularity of the horizon and the Raychaudhuri equation gives a rigidity of the horizon. This rigidity, now extended to the whole past boundary of the universal cover, will then imply that one of the metric function needs to be constant throughout the spacetime, which is a non-generic criterion. 
In the case of collisionless matter, the contradiction arises easily from the rigidity of the past boundary of the fundamental domain and the conservation of the flux of particles as in Theorem 13.2 of \cite{dr:ss}.

The proofs of Theorem \ref{maintheorem} and Theorem \ref{th:vla} rely heavily on these rigidity properties of possible Cauchy horizons which follows once we have shown that Cauchy horizons are necessarily regular. This can be compared with the results of \cite{frw:rtsh} where it was shown that spacetimes which admit a non-degenerate, regular, Cauchy horizon foliated by null closed geodesics have an extra Killing field which is null on the Cauchy horizon. One interesting feature of $T^2$-symmetric spacetimes with $\Lambda >0$ and $r_0 >0$ is that one directly obtains from the Einstein equations that regular Cauchy horizons are necessarily non-degenerate. Moreover, for these models, a Killing field field on the Cauchy horizon could not be one of the Killing fields that generates the $T^2$-symmetry because the orbits would not be null. Therefore, the existence of such a Killing field will imply that these spacetimes are non generic. However, in the case of $T^2$-symmetric spacetimes admitting a regular Cauchy horizon, it is not known whether the null geodesics are closed so the results of \cite{frw:rtsh} do not apply directly after we have obtained the regularity of the Cauchy horizon.

\section{Preliminaries}
\subsection{$T^2$-symmetric spacetimes with spatial topology $T^3$}
A spacetime $(\mathcal{M},g)$ is said to be $T^2$-symmetric if the metric is invariant under the action of the Lie group $T^2$ and the group orbits are spatial. The Lie algebra of $T^2$ is spanned by two commuting Killing fields $X$ and $Y$ everywhere non-vanishing and we may normalise them so that the area element $r$ of the group orbits is given by:
\begin{displaymath}
g(X,X)g(Y,Y)-g(X,Y)^2=r^2.
\end{displaymath}

Moreover, we will require the geometric part of the initial data to be of the form $(\Sigma,g,K)$ with $\Sigma$ diffeomorphic to $T^3$ and $g$ and $K$ as well as the suitable initial data for the matter fields to be invariant under the action of $T^2$. For convenience, we will suppose that the initial data is smooth. The class of initial data for the matter fields is described in appendix \ref{ap:idv}. It can be easily shown that the maximal Cauchy development $(\mathcal{M},g)$ of such initial data has topology $\mathbb{R} \times T^3$ with metric:

\begin{eqnarray} \label{metric}
ds^2=-\Omega^2 [dt^2-d\theta^2]+e^{2U} \left[ dx+Ady+2(G+AH)d\theta \right]^2 \nonumber 
\\+e^{-2U}r^2 \left[dy^2+2 H d\theta \right]^2,
\end{eqnarray}
where all functions depend only on $t$ and $\theta$ and are periodic in the latter.

\paragraph{}The so-called Gowdy spacetimes correspond to the particular cases where the functions $G$ and $H$ vanish everywhere, resulting in the orbits of symmetry being orthogonal to the $t$ and $\theta$ directions\comment{In the vacuum, the Einstein equations imply that if $G$ and $H$ vanish at one point, they vanish everywhere. Therefore initial data can be separated between Gowdy type and T^2 type of initial data. However, for spacetime with matter, such distinction is artificial since a priori, we have no control on the value of G and H from their value on the initial surface.}. This orthogonality implies that there is a natural way of prescribing a Lorentzian metric on the 2-manifold $\mathcal{Q}=\mathcal{M} / T^2$. For general $T^2$-symmetric spacetimes the orbits are not orthogonal to the $t$ and $\theta$ directions but we can still prescribe a metric on the quotient manifold $\mathcal{Q}$ by the rule that the inner product of two vectors on the quotient is equal to the inner product of the unique vectors of the spacetime which project onto them orthogonally to the orbit. 
In the following, the lift of $(\mathcal{Q},g_{\mathcal{Q}})$ to its universal cover will be denoted by $(\mathcal{\widetilde{Q}},\widetilde{g})$ and more generally, we will write $\widetilde{T}$ the lift of $T$ to $\mathcal{\widetilde{Q}}$, for any $T$.

By a change of coordinates of the form $v=\alpha(t+\theta)$, $u=\beta(t-\theta)$, we may rewrite the metric (\ref{metric}) in null coordinates: 

\begin{eqnarray} \label{eq:metn}
ds^2=-\Omega^2 du dv+e^{2U} \left[ dx+Ady+(G+AH)\left[\frac{dv}{\alpha'(v)}-\frac{du}{\beta'(u)}\right] \right]^2 \nonumber 
\\+e^{-2U}r^2 \left[dy+H \left[\frac{dv}{\alpha'(v)}-\frac{du}{\beta'(u)}\right] \right]^2,
\end{eqnarray}
relabelling $\Omega$ suitably. By choosing $\alpha$, $\beta$ appropriately, we can ensure that $\Omega=1$ along a constant $v$ ray or a constant $u$ ray as needed later, but note that by doing so, we will also break the periodicity in $\theta$.

It is easy to see from the form of the metric that any constant $u$ (or $v$) hypersurface is then a null hypersurface, however, in general, a constant $(u,x,y)$ ray will not be a null ray because of the non-orthogonality of the orbits. Let us define the two null vectors which are orthogonal to the orbits by $N_{(u)}$, $N_{(v)}$:

\begin{eqnarray}
N_{(u)}&=& \drond{}{u}+\frac{1}{\beta'}\left(G \drond{}{x}+H \drond{}{y}\right),\\
N_{(v)}&=& \drond{}{v}-\frac{1}{\alpha'}\left(G \drond{}{x}+H \drond{}{y}\right).
\end{eqnarray}

A computation shows that $\Omega^{-2}$ is an affine parameter for $N_{(u)}$, $N_{(v)}$ i.e:

\begin{equation}
\nabla_{\frac{N_{(u)}}{\Omega^2}} \frac{N_{(u)}}{\Omega^2}=0, \quad \nabla_{\frac{N_{(v)}}{\Omega^2}} \frac{N_{(v)}}{\Omega^2}=0.
\end{equation}
The affine lengths of the integral curves of $N_{(u)}$ and $N_{(v)}$ are therefore given respectively by $\int_u \Omega^2 du$ and $\int_v \Omega^2 dv$.

Finally by analogy with \cite{arw:caft2v}, we define the twist quantities:

\begin{eqnarray}
\Gamma&=&2 \frac{(G_t+AH_t)}{\alpha' \beta'}=2 \frac{G_{u}}{\alpha'}+2 \frac{G_v}{\beta'}+2 A \left( \frac{H_{u}}{\alpha'}+\frac{H_v}{\beta'} \right), \label{eq:gamma}\\
\Pi&=&2\frac{H_t}{\alpha' \beta'}= 2\frac{H_{u}}{\alpha'}+2\frac{H_v}{\beta'}.\label{eq:pi}
\end{eqnarray}

\subsection{The Einstein equations in null coordinates}
The Einstein equations give rise to the following system:

Null constraint equations:

\begin{eqnarray}
\partial_v \left( \Omega^{-2} r_v e^{-2U} \right) &=& - \demir \Omega^{-2} e^{2U} A_v^2 - 2r e^{-2U}\Omega^{-2} U_v^2 \nonumber \\
 &-&2 \pi r e^{-2U}\frac{\beta'}{\alpha'}(\rho+P_1-2J_1), \label{ee:rv}\\
\partial_u \left( \Omega^{-2} r_u e^{-2U} \right) &=& - \demir \Omega^{-2} e^{2U} A_u^2 - 2r e^{-2U}\Omega^{-2} U_u^2 \nonumber \\
 &-&2 \pi r e^{-2U}\frac{\alpha'}{\beta'}(\rho+P_1+2J_1). \label{ee:ru}
\end{eqnarray}

Evolution equations:
\begin{eqnarray}
r_{uv}&=&2 \pi r \Omega^2 (\rho-P_1) + \frac{r}{2} \Omega^2 \Lambda +\frac{re^{2U}}{8} \frac{\Gamma^2}{\Omega^2}+\frac{r^{3} e^{-2U}}{8}\frac{\Pi^2}{\Omega^2}, \label{ee:wr}\\
U_{uv} &=& -\demir \left(r_vU_u+r_uU_v \right) +\demirr e^{4U}A_u A_v+\quart \Omega^2 \Lambda \nonumber \\&+&\frac{e^{2U}}{8}\frac{\Gamma^2}{\Omega^2}+\pi \Omega^2 (\rho-P_1+P_2-P_3), \label{ee:wu}\\
A_{uv}&=&-2(A_vU_u+A_uU_v)+\demir (A_ur_v+A_vr_u) \nonumber \\
&+&\frac{r^2e^{-2U}}{4}\frac{\Gamma \Pi}{\Omega^2} +4\pi r \Omega^2 e^{-2U} S_{23}, \label{ee:wa}\\
\partial_u \partial_v \log \Omega&=&-U_u U_v -\quartrr e^{4U} A_u A_v +\demir \left ( r_vU_u+r_u U_v \right ) \nonumber \\&-&\frac{3e^{2U}}{16}\frac{\Gamma^2}{\Omega^2}-\frac{3r^2e^{-2U}}{16}\frac{\Pi^2}{\Omega^2} - \pi \Omega^2(\rho-P_1+P_2+ P_3).\label{ee:wo}
\end{eqnarray}

Auxiliary equations:

\begin{eqnarray} 
\partial_v \left( \frac{\Gamma}{\Omega^2}r e^{2U} \right)=8 \pi r \Omega e^{U} \sqrt{\frac{\beta'}{\alpha'}}(S_{12}-J_2), \\
\partial_u \left( \frac{\Gamma}{\Omega^2}r e^{2U} \right)=8 \pi r \Omega e^{U} \sqrt{\frac{\alpha'}{\beta'}}(S_{12}+J_2), \\
\partial_v \left( \frac{\Pi}{\Omega^2}r^3 e^{-2U} \right)+\frac{\Gamma}{\Omega^2}r e^{2U} A_v=8 \pi r^2 \Omega e^{-U} \sqrt{\frac{\beta'}{\alpha'}}(S_{13}-J_3), \\
\partial_u \left( \frac{\Pi}{\Omega^2}r^3 e^{-2U} \right)+\frac{\Gamma}{\Omega^2}r e^{2U} A_u=8 \pi r^2 \Omega e^{-U} \sqrt{\frac{\beta'}{\alpha'}}(S_{13}+J_3). \label{ee:la}
\end{eqnarray}

$\Gamma$ and $\Pi$ are given by (\ref{eq:gamma}) and (\ref{eq:pi}) and $\rho$, $P_k$, $J_k$, $S_{jk}$ are the components of the energy-momentum tensor in the orthonormal frame:
 
\begin{eqnarray} \label{fr:ort}
E_0&=&\left(-g\left(\drond{}{t},\drond{}{t}\right)\right)^{-1/2}\drond{}{t}, \nonumber \\
E_1&=&\left(-g\left(\drond{}{t},\drond{}{t}\right)\right)^{-1/2}\left(\drond{}{\theta}-G\drond{}{x}-H\drond{}{y}\right),\nonumber \\
E_2&=& e^{-U}\drond{}{x},\nonumber \\
E_3&=& e^U r^{-1} \left(\drond{}{y}-A\drond{}{x} \right). 
\end{eqnarray}

An introduction to collisionless matter can be found in appendix \ref{ap:ve}. Writing the energy-momemtum tensor in an invariant orthornormal frame will be useful later as it is difficult to interpret the energy conditions such as the dominant and the strong energy conditions for the matter in the basis associated with the $(u,v,x,y)$ coordinates.

\paragraph{}
In the following, we shall often use the notation $r_v=\lambda$, $r_u=\nu$. 
\subsection{Global null coordinates on $\mathcal{\widetilde{Q}}$} 
To conclude the preliminaries, let us summarize the existence of $\mathcal{\widetilde{Q}}$ and global null coordinates as follows:

\begin{proposition} \label{prop:gnc} Let $(\mathcal{M},g,f)$ be the maximal Cauchy development of $T^2$-symmetric initial data with non-negative cosmological constant $\Lambda \ge 0$ and (possibly vanishing) Vlasov matter. 
 Let $\mathcal{Q}$ denote the space of group-orbits:

\begin{equation}
\mathcal{Q}=\mathcal{M} / T^2
\end{equation}
and let $\pi_1: \mathcal{M} \rightarrow \mathcal{Q}$ denote the standard projection.

 Then, there exists smooth functions $U^{\mathcal{Q}}$, $A^{\mathcal{Q}}$, $r^{\mathcal{Q}}$, $G^\mathcal{Q}$, $H^\mathcal{Q}$ and $\Omega^{\mathcal{Q}}$ defined on $\mathcal{Q}$, with $r^{\mathcal{Q}}$ and $\Omega^{\mathcal{Q}}$ strictly positive, and a smooth non-negative function $f^{\mathcal{Q}}$ defined on $\mathcal{Q} \times \mathbb{R}^3$ such that the following holds:

\begin{itemize}
\item{There exists a globally defined coordinate system $(t,\theta,x,y)$ covering $\mathcal{M}$ such that $g$ satisfies (\ref{metric}) for some smooth functions $A$, $U$, $r$ and $\Omega$, with $r$ and $\Omega$ stricly positive, and such that all functions are independent of $x$ and $y$ and are periodic with period $1$ in $\theta$.}
\item{$U$, $A$, $r$, $\Omega$, $f$ are the pull-back of $U^{\mathcal{Q}}$, $A^{\mathcal{Q}}$, $r^{\mathcal{Q}}$, $\Omega^{\mathcal{Q}}$ and $f^{\mathcal{Q}}$ by $\pi_1^*$.}
\end{itemize}
Let moreover denote by $\mathcal{\widetilde{Q}}$ the universal cover of $\mathcal{Q}$. Let $\widetilde{U}$, $\widetilde{A}$, $\widetilde{r}$, $\widetilde{G}$ and $\widetilde{H}$ be the lifts of $U^{\mathcal{Q}}$, $A^{\mathcal{Q}}$, $r^{\mathcal{Q}}$, $G^\mathcal{Q}$, $H^\mathcal{Q}$ to $\mathcal{\widetilde{Q}}$ and let $\widetilde{f}$ be the lift of $f^{\mathcal{Q}}$ to $\mathcal{\widetilde{Q}} \times \mathbb{R}^3$. Let finally $\alpha$ and $\beta$ be two smooth functions defined on $\mathbb{R}$ and such that $\alpha' >0$ and $\beta' >0$. Then the following holds:
\begin{itemize}
\item{There exists global null future-directed coordinates $u$ and $v$ on  $\mathcal{\widetilde{Q}}$ and a smooth strictly positive function $\widetilde{\Omega}$ such that $\widetilde{\Omega}^2\alpha'(v)\beta'(u)$ is the lift of $\Omega^2$ to $\mathcal{\widetilde{Q}}$,}
\item{$\widetilde{U}$, $\widetilde{A}$, $\widetilde{r}$, $\widetilde{G}$, $\widetilde{H}$, $\widetilde{\Omega}$, $\alpha$, $\beta$ and $\widetilde{f}$ satisfy the system (\ref{ee:rv})-(\ref{ee:la}), where in the case of non-vanishing Vlasov matter, $\rho$, $P_k$, $S_{jk}$ are defined by (\ref{em:rho})-(\ref{em:S}) and all functions in the equations should be replaced by their tilde versions.}

\end{itemize}

\end{proposition}
In the above proposition, we have assumed smoothness of the initial data for convenience but one could easily obtain statements for initial data lying in a lower class of differentiability.
\begin{proof}
The proof follows by standard arguments as found in \cite{bicm:gft2, arw:caft2v, ci:t2pcc} and a change of coordinates of the type $v=\alpha(t+\theta)$, $u=\beta(t-\theta)$.
\end{proof}
The above proposition shows the existence of a global null coordinate system on $\mathcal{\widetilde{Q}}$, but note that we will allow ourselves to move to other global null coordinate systems, by rescaling $u$ and $v$, as for instance in Proposition \ref{prop:id}. In other words, by choosing $\alpha$ and $\beta$ appropriately, we will simplify our analysis. The rules which dictate the change of coordinates are described in detail in appendix \ref{ap:gnc}.

In the rest of this article, we will, by an abuse of notation, drop the tilde on the functions defined on $\mathcal{\widetilde{Q}}$.
\section{The theorems}

\begin{theorem} \label{maintheorem}
Let $(\mathcal{M},g)$ be the maximal Cauchy development of vacuum $T^2$-symmetric initial data with positive cosmological constant $\Lambda>0$ as described above and suppose that:
\begin{enumerate}
\item{There exists a maximal past-directed causal geodesic $\gamma(t)$ such that $r(t) \rightarrow r_0 > 0$.}
\item{The Killing fields generating the $T^2$-symmetry cannot be chosen so that they are mutually orthogonal on the initial Cauchy surface.}  \label{as:kf}
\end{enumerate}
Then $(\mathcal{M},g)$ is past inextendible as a $C^2$ Lorentzian manifold.
\end{theorem}

Assumption \ref{as:kf} is equivalent to the statement that the function $A$ as defined in (\ref{metric}) is not constant in $\mathcal{M}$.

\begin{theorem} \label{th:vla}
Let $(\mathcal{M},g)$ be the maximal Cauchy development of $T^2$-symmetric initial data with positive cosmological constant $\Lambda>0$ and collisionless matter as described above and suppose that:
\begin{enumerate}
\item{There exists a maximal past directed causal geodesic $\gamma(t)$ such that $r(t) \rightarrow r_0 > 0$.}
\item{The Vlasov field $f$ does not vanish everywhere.}

\end{enumerate}
Then $(\mathcal{M},g)$ is past inextendible as a $C^2$ Lorentzian manifold.
\end{theorem}

Note that for any maximal Cauchy development of $T^2$-symmetric data with $\Lambda \ge 0$ and collisionless matter, it is easy to see that $r$ tends to the same limit $r_0$ for all maximal past directed geodesics.

Since the cases where $\Lambda=0$ were treated in \cite{dr:t2} and since the methods of \cite{dr:t2} may also be applied in the cases where $\Lambda > 0$ and $r(t) \rightarrow 0$ along past directed causal geodesics, we immediately obtain from Theorem \ref{th:vla} and Theorem 4.1 of \cite{dr:t2}:
\begin{corollary} \label{completevlasov}
Let $(\mathcal{M},g)$ be the maximal Cauchy development of $T^2$-symmetric data with non-negative cosmological constant and collisionless matter and let $f: \mathcal{P} \rightarrow \mathbb{R}$ denote the Vlasov field and suppose that:
\begin{enumerate}
\item{There exists a constant $\delta>0$ such that for any open $\mathcal{U} \subset \mathcal{P} \cap \pi^{-1}(\Sigma)$ we have that f does not vanish identically on $\mathcal{U} \cap \{p:g(p,X)^2+g(p,Y)^2 < \delta \}$.} 
\end{enumerate}
Then $(\mathcal{M},g)$ is inextendible as a $C^2$ Lorentzian manifold.
\end{corollary}

Finally, since future inextendibility holds for $T^2$-symmetric spacetimes with non-negative cosmological constant and collisionless matter in view of the results of \cite{dr:iecs}, we can replace \emph{past inextendible} by \emph{past and future inextendible} in Theorem  \ref{maintheorem}, Theorem \ref{th:vla} and Corollary \ref{completevlasov}.

\section{Initial data for the estimates} \label{se:id}
In this section, we will show how we can construct appropriate initial data to perform estimates of the type found in \cite{arw:caft2v} in bounded null coordinates. We therefore want to establish the existence of a bounded null coordinate system $(u,v)$ such that along a certain null ray $v=v_1$, $A$, $U$ and $r$ as well as their first derivatives are bounded. First, we will need the following:

\subsection{Extendibility of the Killing vector fields} 
\begin{lemma}\label{exkf}
Let $(\mathcal{M},g)$ be the past maximal Cauchy development of $T^2$-sym\-metric initial data with positive cosmological constant, in the vacuum or with Vlasov matter, as in Theorem \ref{maintheorem} and \ref{th:vla}. Suppose that $r_0 >0$ and that $(\mathcal{M},g)$ is extendible as a Lorentzian manifold with $C^2$ metric. Let $\gamma$ be a causal geodesic leaving $(\mathcal{M},g)$ and $p$ be the intersection of $\gamma$ with the past boundary of $\mathcal{M}$. Then $r$, $U$ and $A$ admit a $C^1$ extension along $\gamma$ to $p$.
\end{lemma}
\begin{proof}
We know from \cite{dr:iecs} that the Killing vector fields generating the surface of symmetry must have $C^1$ extensions to the Cauchy horizon in any $C^2$ extension of the spacetime. Suppose that $\gamma$ is a causal geodesic leaving $(\mathcal{M},g)$ and let $p$ be the intersection of $\gamma$ with the past boundary of $\mathcal{M}$. We then have that $g(X,X)$, $g(X,Y)$, $g(Y,Y)$ are $C^1$ functions along $\gamma$ and in particular, they are bounded along $\gamma$ until $p$. From (\ref{metric}), we have:

\begin{eqnarray}
g(X,X)&=& e^{2U}, \label{eq:gxx}\\
g(X,Y)&=& e^{2U}A, \label{eq:gxy}\\
g(Y,Y)&=& r^2e^{-2U}+A^2e^{2U}. \label{eq:gyy}
\end{eqnarray}
From (\ref{eq:gyy}), we see that $e^{-2U}$ is bounded above since $r_0>0$, which implies that $U$ is bounded below. By (\ref{eq:gxx}) and (\ref{eq:gxy}), this implies that $r$, $U$ and $A$ are at least $C^1$ along $\gamma$.

\end{proof}

Using this lemma, we may then reduce the issue of inextendibility to this of orbits-orthogonal null geodesic inextendibility.

\subsection{Reduction to orbits-orthogonal null geodesic inexten\-dibility} \label{se:ri}
We will adapt the method of proposition 13.1 of \cite{dr:ss} to our geometry in order to prove the following lemma:

\begin{lemma}
Let $(\mathcal{M},g)$ be the past maximal Cauchy development of $T^2$-sym\-metric initial data with positive cosmological constant, in the vacuum or with Vlasov matter, as in Theorem \ref{maintheorem} and \ref{th:vla}. Suppose that $r_0 >0$ and that $(\mathcal{M},g)$ is extendible as a Lorentzian manifold with $C^2$ metric. Then there exists a null line, orthogonal to the orbits of symmetry, which leaves $(\mathcal{M},g)$ and enters a $C^2$ extension of $(\mathcal{M},g)$.
\end{lemma}

\begin{proof}

 Note that for any geodesic, the following quantities (conservation of angular momentum) are conserved:

\begin{align}
J_x=& e^{2U} \left[ \dot{x}+A \dot{y}+(G+AH)\left(\frac{\dot{v}}{\alpha'}-\frac{\dot{u}}{\beta'} \right) \right], \\
J_y=& r^2 e^{-2U}\left[\dot{y}+H\left(\frac{\dot{v}}{\alpha'}-\frac{\dot{u}}{\beta'} \right)\right]+AJ_x.
\end{align}

Let $\mathcal{H^+}$ denote the past boundary of $\mathcal{M}$ in an extension $\mathcal{M}'$ and let $p \in \mathcal{H}^+$ be such that there exists no null geodesic orthogonal to the orbits leaving the spacetime in a neighbourhood of $p$.

Following \cite{dr:ss}, we can find a sequence of regular points $p_i \in \mathcal{H}^+$ converging to $p$ on $\mathcal{H}^+$, and planes $O_i$, $T_i$ such that $O_i$ and $T_i$ are respectively, the planes orthogonal and tangent to the orbits of symmetry. The planes $O_i$ can also be regarded as the set of vectors with vanishing angular momentum $J_x$, $J_y$. Conservation of angular momentum implies that the planes $O_i$ are null and their null generator $K_i$ is necessarily tangential to $H_{p_i}^+$. We can then extract a subsequence $O_i$ converging to a necessarily null plane $O$ at $p$. We may then draw a convergent subsequence $T_i$ converging to $T$. Since $T_i$ and $O_i$ are orthogonal, $T_i$ is also null and there exists a null vector $K \in O \cap T$.

The achronality of $\mathcal{H}^+$ at $p$ implies the existence of timelike geodesics entering $\mathcal{M}$ at $p$. If $\gamma$ is such a geodesic, $g(K, \dot{\gamma}) \neq 0$. Let $K_j$ be a sequence of vectors tangential to the orbits of symmetry along $\gamma$ and converging to $K$. The Cauchy-Schwarz inequality implies:

\begin{equation}
g(K_j, \dot{\gamma}) \le g(K_j,K_j)^{1/2}(e^{-2U}J_x^2+r^{-2}e^{2U}(J_y-A J_x)^2).
\end{equation}
Since we have previously shown that $U$ and $A$ are bounded and $J_x$ and $J_y$ are constant along $\gamma$, as $K_j$ goes to $K$, the right hand side goes to zero which is a contradiction.
\end{proof}

In order to provide the initial data necessary to perform the estimates of the next sections, we first need to understand some basic notions concerning the local and the global geometry of $T^2$-symmetric spacetimes. First, we will need the following:

\subsection{Monotonicty of r}
\begin{lemma}
Let $(\mathcal{M},g)$ be the past maximal Cauchy development of $T^2$-sym\-metric initial data with positive cosmological constant, in the vacuum or with Vlasov matter, as in Theorem \ref{maintheorem} and \ref{th:vla}. Then the gradient of $r$ is timelike, which in double null coordinates means that:

\begin{equation}
\lambda  \nu >0
\end{equation}
and by a choice of orientation that:

\begin{equation}
\lambda>0, \quad  \nu >0.
\end{equation}

\end{lemma}
\begin{proof}
It is easy to see that propostion 3.1 of \cite{ar:cmc2ds} holds for $T^2$-symmetric spacetimes with non-negative cosmological data and collisionless matter. It follows that either $\lambda \nu >0$ or the spacetime is flat. Since the term containing the cosmological constant on the right hand side of (\ref{ee:wr}) ensures that the latter case does not occur, we have $\lambda  \nu >0$. Choosing the time orientation such that the future corresponds to the expanding direction, we can assume $\lambda >0$, $\nu>0$.
\end{proof}

\subsection{Global structure of $\mathcal{\widetilde{Q}}$}
In this section, we describe some global geometric features of $\mathcal{\widetilde{Q}}$.

\begin{lemma} \label{lem:gg} Let $(\mathcal{M},g)$ be the past maximal Cauchy development of $T^2$-sym\-metric initial data with positive cosmological constant, in the vacuum or with Vlasov matter, as in Theorem \ref{maintheorem} or \ref{th:vla}. Suppose that $r_0 >0$. Then all integral curves of $N_{(u)}$ and $N_{(v)}$ are incomplete in the past direction.
\end{lemma}
\begin{proof}
Since the dominant energy condition holds for Vlasov matter, we have $\rho-P_1 \ge 0$. Therefore all terms on the right-hand side of equation (\ref{ee:wr}) are non-negative and we have for all $(u,v)$ and fixed $u'$,

\begin{equation}
\int^{u'}_{u}\Lambda \Omega^2 r du \le \int^{u'}_{u} 2r_{uv} \le 2\lambda(u_1,v)
\end{equation}
which implies, using the lower bound $r \ge r_0 >0$, that:

\begin{equation}
\int^{u'}_{u}\Omega^2  du \le \frac{2\lambda(u',v)}{\Lambda r_0}
\end{equation}
and similar inequalities hold for constant $u$ null curves. Since $\Omega^2$ is an affine parameter for the integral curves of  $N_{(u)}$ and $N_{(v)}$, we have proved that their affine length is bounded.
\end{proof} 

We may then draw a Penrose diagram for $\mathcal{\widetilde{Q}}$. Assuming $r_0 > 0$, it is easy to see that the range of the coordinates $(t, \theta, x, y)$ as in (\ref{metric}) is given by $\mathbb{R} \times [0,1]^3$, using the classical energy estimates for $T^2$-symmetric spacetimes as found in \cite{bicm:gft2, arw:caft2v, ci:t2pcc}. Taking a parametrisation $u=\alpha(t-\theta)$, $v=\beta(t+\theta)$ such that $\Omega=1$ along a constant $u$ and a constant $v$ null curves, we can represent the Penrose diagram of $\mathcal{\widetilde{Q}}$ by (see \cite{dr:ss} for example):

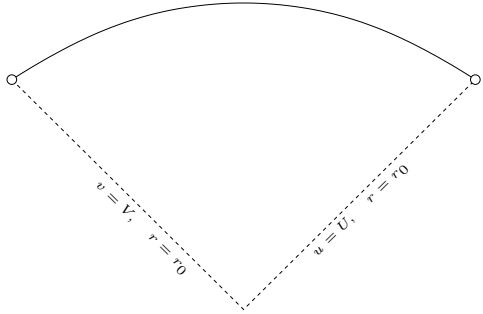
\[ \input{penrose_diagram.pstex_t} \]

\paragraph{}

\subsection{The initial data} \label{ni:c1}

We are now ready to build a null coordinate system and initial data tailored for the estimates of the next section.

\begin{proposition} \label{prop:id}
Let $(\mathcal{M},g)$ be the past maximal Cauchy development of $T^2$-symmetric initial data with positive cosmological constant, in the vacuum or with Vlasov matter, as in Theorem \ref{maintheorem} or \ref{th:vla}. Suppose that $r_0 >0$ and that there exists an integral curve of $N_{(u)}$ which leaves $(\mathcal{M},g)$ and enters a $C^2$ extension of $(\mathcal{M},g)$. Then, there exists a double null bounded coordinate system $(u,v)$ covering  $\mathcal{\widetilde{Q}}$, with null lines $v=v_1$, $v=v_0$, $u=u_1$, and a constant $\Omega_0>0$ such that the following holds:

\begin{enumerate}
\item{Along $v=v_1$, $r$, $U$, $A$ and their first derivatives may be continuously extended to $u=U$.}
\item{The null curve $v=v_0$ is the image by the deck transformation of the null curve $v=v_1$.}
\item{Moreover, the following holds:
\begin{eqnarray}
\Omega(u,v_1)&=&\Omega_0, \\
\sup_{u \in (U,u_1]}\left(\nu(u,v_1) \right)\left(\frac{v_1-v_0}{r_0}+\frac{3}{\sqrt{r_0}}\right) &=&1,  \label{eq:supnu}\\
\lambda(u_1,v)&=&1.
\end{eqnarray}
}
\end{enumerate}
\end{proposition}
\begin{proof}
From Proposition \ref{prop:gnc}, there exist global null coordinates $(u,v)$ on $\mathcal{\widetilde{Q}}$. In order to build null coordinates such that the estimates of the proposition hold, we will rescale $(u,v)$ and then exploit the regularity of $\nu$ along the null curve leaving the spacetime to construct another null coordinate system, $(\tilde{\tilde{u}},\tilde{\tilde{v}})$, satisfying the requirements of the proposition.

Let thus $\gamma$ be an integral curve of $N_{(u)}$ leaving the spacetime and suppose that in the original coordinate system $(u,v)$, $\gamma$ is given by $v=v_1$. Fix moreover a null line $u=u_1$ . Let $\mathcal{T}$ be $\mathcal{\widetilde{Q}} \cap \{(u,v) /\quad u \le u_1,  v \le v_1\}$. Let $v=v_0$ be the image by the deck transformation of the null line $v=v_1$. One may visualise $\mathcal{T}$ and the null curves $v=v_1$, $v=v_0$ and $u=u_1$,  in a Penrose diagram of $\mathcal{\widetilde{Q}}$:
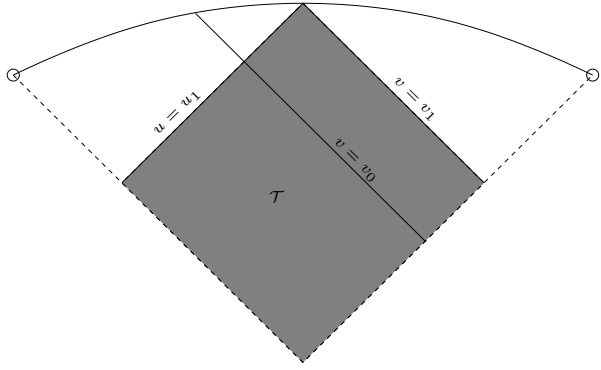
\[ \input{penrose_diagram_integration_idtip.pstex_t} \]

We define a new coordinate system $(\tilde{u}, \tilde{v})$ as follows. First define $\tilde{v}$ by:

\begin{equation} 
\tilde{v}(v)=v_1-\int_v^{v_1}\lambda(u_1,v')dv'=v_1-r(u_1,v_1)+r(u_1,v).
\end{equation}
It follows immediately from the bound on $r$ that $\tilde{v}$ is bounded and thus take values in $(V,v_1]$ for some finite $V$. Using $\tilde{v}$ as a replacement for $v$, one obtain a coordinate system $(u,\tilde{v})$ for which $\tilde{\lambda}=r_{\tilde{v}}$ satisfies:

\begin{equation} \label{eq:lambdau1}
\tilde{\lambda}(u_1,\tilde{v})=1.
\end{equation}
In the following, we drop the tilde on $\tilde{v}$.

Let $\Omega_0$ be a stricly positive constant. Define $\tilde{u}$ by :

\begin{equation}
\tilde{u}=u_1- \int^{u_1}_u\frac{\Omega^2(u',v_1)}{\Omega_0^2}du'.
\end{equation}
From lemma \ref{lem:gg}, the affine length of $v=v_1$ is finite. Thus, $\tilde{u}$ is bounded from below and takes value in $(U,u_1]$ for some finite $U$. Moreover, $(\tilde{u},v)$ defines a new coordinate system for which $\tilde{\Omega}^2=-2 g_{\tilde{u}v}$ satisfies:

\begin{equation}
\tilde{\Omega}^2(u,v_1)=\Omega_0^2.
\end{equation}
Note that since $\tilde{u}$ is a function of $u$ only, the change of coordinate does not affect equation (\ref{eq:lambdau1}). Note also that since $r$ is a purely geometric quantity, it is invariant under a change of coordinates on $\mathcal{\widetilde{Q}}$, and therefore, that $r_0$ is invariant.

Since $\Omega$ is constant along $v=v_1$, it then follows that $N_{(\tilde{u})}$ is parallely transported along $v=v_1$. From lemma \ref{exkf}, $r$ is a $C^1$ function along $v=v_1$ up to $u=U$. In particular, $\tilde{\nu}=r_{\tilde{u}}$ is bounded along $v=v_1$. Let $A$ denote $\sup(\nu (u,v_1))\left(\frac{v_1-v_0}{r_0}+\frac{3}{\sqrt{r_0}}\right)$. We define $\tilde{\tilde{u}}$ by :

\begin{equation}
\tilde{\tilde{u}}= A \tilde{u}.
\end{equation}
In the coordinate system given by $(\tilde{\tilde{u}},v)$, $\tilde{\tilde{\Omega}}(\tilde{\tilde{u}},v_1)$ is also constant, the constant being given by $\frac{\Omega_0}{ \sqrt{A}}$. Thus,  $N_{(\tilde{\tilde{u}})}$ is also parallely transported along $v=v_1$ and it follows from lemma \ref{exkf} that $r$, $A$, $U$ and their first derivatives admit continuous extension along $v=v_1$ to $u=U$.

Moreover, we have:

\begin{equation}
\tilde{\tilde{\nu}}=\frac{\tilde{\nu}}{A}.
\end{equation}
In particular, $\tilde{\tilde{\nu}}(\tilde{\tilde{u}},v_1) \left(\frac{v_1-v_0}{r_0}+\frac{3}{\sqrt{r_0}}\right)\le 1$ and:

\begin{equation}
 \sup\left(\tilde{\tilde{\nu}}(\tilde{\tilde{u}},v_1) \left(\frac{v_1-v_0}{r_0}+\frac{3}{\sqrt{r_0}}\right)\right)=1.
\end{equation}

Thus, $(\tilde{ \tilde{u}},v)$ satisfies the requirements of the proposition.

\end{proof}

\section{Estimates in a fundamental domain of the universal cover} \label{se:es}
In this section, we will assume that Proposition \ref{prop:id} holds in order to derive energy estimates in a fundamental domain of $\mathcal{\widetilde{Q}}$. Let therefore $(u,v)$ be the bounded null coordinate system of Proposition \ref{prop:id}, such that $\mathcal{T}=(U,u_1] \times (V,v_1]$ and $v=v_1$ is an integral curve of $N_{(u)}$ leaving the spacetime. Let $v=v_0$ be the image by the deck transformation of $v=v_1$ and define $\mathcal{F}$ by:

\begin{equation}
\mathcal{F}=(U,u_1] \times [v_0,v_1].
\end{equation}
We may represent $\mathcal{F}$ in a Penrose diagram:

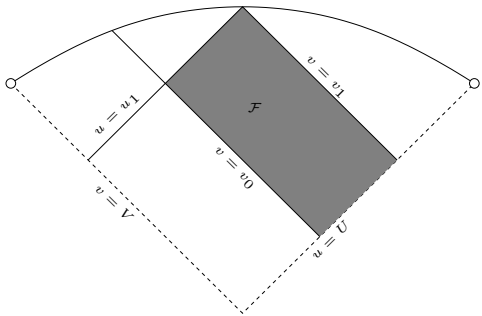
\[ \input{penrose_diagram_F.pstex_t} \]

 We will perform estimates in $\mathcal{F}$, using the bounds we previously obtained along $v=v_1$ in section \ref{se:id} and the bounds coming from the compactness of the intersection of $u=u_1$ with $\mathcal{F}$.

\subsection{Uniform bounds on $\nu$, $\lambda$ in a fundamental domain} \label{se:ubln}
From (\ref{ee:wr}), it follows that $r_{uv} \ge 0$ and therefore $\nu$ and $\lambda$ are bounded above by their values on the intersection of $u=u_1$ with $\mathcal{F}$ and their values on $v=v_1$.

\subsection{Uniform bounds on $U_u$, $U_v$, $A_u$, $A_v$} \label{es:AU}
The method follows \cite{bicm:gft2, arw:caft2v, ci:t2pcc} but the final Gronwall type argument needs to be modified to be adapted to the null bounded coordinate system and the facts that $\Lambda >0$ and that matter fields may be present.

We define the quantities $G$ and $H$ by:

\begin{eqnarray}
G&=r(U_v^2+U_u^2)+\frac{e^{4U}}{4r}(A_v^2+A_u^2), \\
H&=r(U_v^2-U_u^2)+\frac{e^{4U}}{4r}(A_v^2-A_u^2). 
\end{eqnarray}

The positive quantities $G+H$, $G-H$ satisfy:

\begin{align}
\partial_u(G+H)&=-2U_vU_u \lambda + \lambda \frac{e^{4U}}{2r^2}A_u A_v +r U_v e^{2U}\frac{\Gamma^2}{2\Omega^2}+ \frac{rA_ve^{2U}}{4}\frac{\Gamma \Pi}{ \Omega^2}\nonumber \\ &+rU_v \Omega^2 \Lambda +e^{2U}A_v 4\pi \Omega^2 S_{23}+ r U_v 4 \pi \Omega^2 (\rho-P_1+P_2-P_3), \label{eq:ham} \\ 
\partial_v(G-H)&=-2U_vU_u \nu + \nu \frac{e^{4U}}{2r^2}A_u A_v+ r U_v e^{2U}\frac{\Gamma^2}{2\Omega^2} + \frac{rA_ue^{2U}}{4} \frac{\Gamma \Pi}{\Omega^2} \nonumber\\ &+rU_u \Omega^2 \Lambda +e^{2U}A_u 4\pi \Omega^2 S_{23}+ r U_u 4 \pi\Omega^2 (\rho-P_1+P_2-P_3).  \label{eq:lag} 
\end{align}

We can integrate (\ref{eq:ham}) and (\ref{eq:lag}) along constant $v$ and $u$ curves:

\[ \input{penrose_diagram_integration_null_rays.pstex_t} \]

We obtain:

\begin{align}
[G+H](u,v)&=[G+H](u_1,v)+\int^{u_1}_{u} \left( 2U_uU_v \lambda - \lambda \frac{e^{4U}}{2r^2} A_u A_v \nonumber \right)\\
&-\int^{u_1}_{u}\left(r U_v e^{2U}\frac{\Gamma^2}{2\Omega^2}+\frac{e^{2U}}{4}rA_v \frac{\Gamma \Pi}{\Omega^2}+rU_v \Omega^2 \Lambda \right) \nonumber \\
&-\int^{u_1}_{u}\left(e^{2U}A_v 4\pi\Omega^2 S_{23}+ r U_v 4\pi \Omega^2(\rho-P_1+P_2-P_3)\right), \\
[G-H](u,v)&=[G-H](u,v_1)+\int^{v_1}_{v}\left( 2U_uU_v \nu - \nu \frac{e^{4U}}{2r^2} A_u A_v \right) \nonumber \\
&-\int^{v_1}_{v}\left(r U_u e^{2U}\frac{\Gamma^2}{2\Omega^2}+\frac{e^{2U}}{4}rA_u \frac{\Gamma \Pi}{\Omega^2}+rU_u \Omega^2 \Lambda \right) \nonumber \\
&-\int^{v_1}_{v}\left(e^{2U}A_u4\pi \Omega^2 S_{23}+ r U_u 4\pi \Omega^2(\rho-P_1+P_2-P_3)\right).
\end{align}

Note that from the definition of the Vlasov matter\footnote{See appendix \ref{ap:ve}.} we have:

\begin{equation}
\rho\ge P_1+P_2+P_3 , \quad P_2+P_3 \ge 2|S_{23}|, \quad P_i \ge 0, \forall i=1,2,3.
\end{equation}
From the first inequality we obtain:

\begin{equation}
(\rho -P_1+P_2-P_3) \le 2(\rho-P_1).
\end{equation}

Using these inequalities as well as $2|U_u U_v|+\frac{e^{4U}}{2r^2} |A_u A_v| \le \frac{G}{r}$, $|U_i|\le (G/r)^{1/2}$, $e^{2U}|A_i| \le 2\sqrt{rG}$ for $i=u,v$, we obtain:

\begin{align}
[G+H](u,v)& \le[G+H](u_1,v)+\int^{u_1}_{u}\lambda \frac{G}{r} +\int^{u_1}_{u} r \left(\frac{G}{r} \right)^{1/2} e^{2U}\frac{\Gamma^2}{2\Omega^2} \nonumber\\
&+\int^{u_1}_{u}  \frac{r |\Gamma \Pi|}{4 \Omega^2}2 \sqrt{rG}+ \sqrt{rG}\Omega^2 \Lambda \nonumber \\
&+\int^{u_1}_{u}2\sqrt{rG}4\pi \Omega^2 S_{23}+ \sqrt{rG} 8\pi \Omega^2 (\rho-P_1), \\
& \le[G+H](u_1,v)+\int^{u_1}_{u}\lambda \frac{G}{r} +\int^{u_1}_{u} r \left(\frac{G}{r} \right)^{1/2} e^{2U}\frac{\Gamma^2}{2\Omega^2} \nonumber\\
&+\int^{u_1}_{u}  \frac{r |\Gamma \Pi|}{4 \Omega^2}2 \sqrt{rG}+ \sqrt{rG}\Omega^2 \Lambda+\sqrt{rG}12 \pi \Omega^2 (\rho-P_1).
\end{align}
Using the inequality, $2ab \le a^2+b^2$ to estimate the product of the twist quantities, we obtain:

\begin{align}
[G+H](u,v)& \le[G+H](u_1,v)+\int^{u_1}_{u}\lambda \frac{G}{r} +\int^{u_1}_{u} r \left(\frac{G}{r} \right)^{1/2} e^{2U}\frac{\Gamma^2}{2\Omega^2} \nonumber\\
&+\int^{u_1}_{u}2 \left(\frac{G}{r}\right)^{1/2} \left( \frac{\Gamma^2 e^{2U}r}{8\Omega^2}+\frac{\Pi^2 e^{-2U}r^3}{8\Omega^2} \right) \nonumber\\
 &+\int^{u_1}_{u}\left( \sqrt{rG}\Omega^2 \Lambda +\sqrt{rG}12 \pi \Omega^2 (\rho-P_1) \right).
\end{align}
We then use equation (\ref{ee:wr}) to estimate all the matter terms as well as the terms containing the twist quantities and the cosmological constant:
\begin{align}
[G+H](u,v)& \le[G+H](u_1,v)+\int^{u_1}_{u}\left(\lambda \frac{G}{r} +\frac{6}{\sqrt{r}}r_{uv}\sqrt{G}\right).
\end{align}

Thus, using the compactness of $\{u_1\} \times [v_0,v_1]$, there exists a positive constant $C$ depending on the value of the metric functions on $\mathcal{F} \cap \{u=u_1 \}$ such that:

\begin{align}
[G+H](u,v)& \le C+\int^{u_1}_{u}\left(\lambda \frac{G}{r} +\frac{6}{\sqrt{r}}r_{uv}\sqrt{G}\right).\label{ham:int} 
\end{align}

Similarly, we have:

\begin{align} \label{ineq:lagint}
[G-H](u,v) \le& C'+ \int^{v_1}_{v}\left( \nu \frac{G}{r} + \frac{6}{\sqrt{r}}r_{uv}\sqrt{G} \right), 
\end{align}
for some positive constant $C'$.

To close the estimates, one would like to apply a Gronwall type argument to the inequalities (\ref{ham:int}) and (\ref{ineq:lagint}). One cannot do this directly, as the right-hand sides of (\ref{ham:int}) and (\ref{ineq:lagint}) depends on $G$ and not on, respectively, $G+H$ and $G-H$. One is therefore tempted to add the two inequalities, in order to obtain $G$ on the left-hand side. Using the estimate $\sqrt{G} \le \frac{G+1}{2}$, we would obtain an inequality of the form:

\begin{equation}
G(u,v) \le C+\int^{u_1}_{u}\left(\left(D+E r_{uv}\right)G\right)(u',v)du'+\int^{v_1}_{v}\left(\left(D+ E r_{uv}\right)G\right)(u,v')dv'.
\end{equation}
Note that it is not possible a priori to apply directly Gronwall's inequality to the above inequality, for instance by considering $v$ to be fixed and applying a Gronwall argument as for a function of the variable $u$ only, as by doing so, one would then need to estimate a volume integral of $G$. In the vacuum case with no cosmological constant, there are no terms containing $r_{uv}$ on the right-hand sides of (\ref{ham:int}) and (\ref{ineq:lagint}). One can then take a supremum over one variable and apply a Gronwall argument to conclude, see \cite{bicm:gft2}. However, when terms containing $r_{uv}$ appear, one cannot take a supremum anymore, as it is not known, a priori, that the integral of the supremum of $r_{uv}$ is bounded. 

This is why we will not apply the estimate $\sqrt{G} \le \frac{G+1}{2}$ to the inequality (\ref{ham:int}). On the other hand, because of the identity (\ref{eq:supnu}), we will not need to be particularly careful with the terms arising from (\ref{ineq:lagint}) in order to close the estimates. Thus, we may apply the estimate $\sqrt{G} \le \frac{G+1}{2}$ in the inequality (\ref{ineq:lagint}):

\begin{align}
[G-H](u,v) \le& C'+ \int^{v_1}_{v}\left( \nu \frac{G}{r} + \frac{3}{\sqrt{r}}r_{uv}(G+1) \right), \\
[G-H](u,v) \le& C + \int^{v_1}_{v}\left( \left(\frac{\nu}{r}+3\frac{r_{uv}}{\sqrt{r}} \right)G \right)(u,v') dv', \label{lag:int}
\end{align}
for some positive constant $C$ depending on the bounds of the metric functions along $v_1$ coming from Proposition \ref{prop:id}.

Adding (\ref{ham:int}) and (\ref{lag:int}), we obtain:

\begin{align}
G(u,v) \le C &+ \frac{1}{2}\int^{u_1}_{u}\left(\lambda \frac{G}{r} +\frac{6}{\sqrt{r}}r_{uv}\sqrt{G}\right)(u',v)du'\nonumber \\&+\frac{1}{2}\int^{v_1}_{v}\left( \left( \frac{\nu}{r}+3\frac{r_{uv}}{\sqrt{r}}\right)G\right)(u,v')dv'
\end{align}
and since $v \ge v_0$ for all $(u,v) \in \mathcal{F}$, we have:

\begin{align}
\le  C &+ \frac{1}{2}\int^{u_1}_{u}\left(\lambda \frac{G}{r} +\frac{6}{\sqrt{r}}r_{uv}\sqrt{G}\right)(u',v)du'\nonumber \\&+\frac{1}{2}\int^{v_1}_{v_0}\left( \left( \frac{\nu}{r}+3\frac{r_{uv}}{\sqrt{r}} \right)G\right)(u,v')dv'.
\end{align}

Since $[v_0,v_1]$ is compact and $G$ continuous, there exists, for every $u' \in [u,u_1]$, a $v_m(u') \in [v_0,v_1]$ such that $G(u',v_m(u'))=\sup_{[v_0,v_1]}{G(u',.)}$ and we can define $F(u')$ by $F(u')=\sup_{[v_0,v_1]}{G(u',.)}=G(u',v_m(u'))$. We have:

\begin{align}
G(u,v) \le& C + \frac{1}{2}\int^{u_1}_{u}\left(\lambda \frac{G}{r} +\frac{6}{\sqrt{r}}r_{uv}\sqrt{G}\right)(u',v)du'\nonumber \\
 &+\frac{1}{2}F(u)\int^{v_1}_{v_0}\left(\frac{\nu}{r}+3\frac{r_{uv}}{\sqrt{r}} \right)(u,v')dv', \\
\le& C + \frac{1}{2}\int^{u_1}_{u}\left(\lambda \frac{G}{r} +\frac{6}{\sqrt{r}}r_{uv}\sqrt{G}\right)(u',v)du'\nonumber \\&+\frac{1}{2}F(u)\left(\sup_{\mathcal{T}}{\nu}\right)\left(\frac{v_1-v_0}{r_0}+\frac{3}{\sqrt{r_0}} \right). \label{ineq:int}
\end{align}

From equation (\ref{eq:supnu}) of Proposition \ref{prop:id} and the fact that $r_{uv} \le 0$ coming from equation (\ref{ee:wr}), we have that $\left(\sup_{\mathcal{T}}{\nu}\right)\left(\frac{v_1-v_0}{r_0}+\frac{3}{\sqrt{r_0}} \right) = 1$. Thus, we obtain:

\begin{align}
G(u,v) \le  &C+ \frac{1}{2}\int^{u_1}_{u}\left(\lambda \frac{G}{r} +\frac{6}{\sqrt{r}}r_{uv}\sqrt{G}\right)(u',v)du'\nonumber \\&+\frac{1}{2}G\left(u,v_m(u)\right). \label{ineq:key}
\end{align}

We can evaluate (\ref{ineq:key}) at the point $(u,v_m(u))$:

\begin{align}
F(u)=G(u,v_m(u)) \le C &+ \frac{1}{2}\int^{u_1}_{u}\left(\lambda \frac{G}{r} +\frac{6}{\sqrt{r}}r_{uv}\sqrt{G}\right)(u',v_m(u))du'\nonumber \\&+\frac{1}{2}F(u).
\end{align}

It follows that:

\begin{equation}
F(u) \le 2C+ \int^{u_1}_{u}\left(\lambda \frac{G}{r} +\frac{6}{\sqrt{r}}r_{uv}\sqrt{G}\right)(u',v_m(u))du'
\end{equation}
and therefore:

\begin{align}
F(u) \le & 2 C + \int^{u_1}_{u}\frac{\lambda(u',v_m(u))}{r_0}F(u')du' \nonumber \\
&+\sqrt{\sup_{[u,u_1]}{F}}\int^{u_1}_{u}\frac{6r_{uv}(u',v_m(u))}{\sqrt{r_0}} du',\nonumber \\
\le & 2C+ B \int^{u_1}_{u}F(u') du'+D\sqrt{\sup_{[u,u_1]}{F}},
\end{align}
for some constant $B$ and $D$ depending on the uniform bound on $\lambda$ obtained in section \ref{se:ubln}.

Apply now an inequality of Gronwall type to obtain:

\begin{equation}
F(u) \le 2C+D\sqrt{\sup_{[u,u_1]}{F}}+\int^{u_1}_u  \left(2C+D\sqrt{\sup_{[u',u_1]}{F}} \right)B \exp(B (u'-u))du'.
\end{equation}
Note the trivial fact that:

\begin{equation}
\sup_{u' \in [u,u_1]}{\left (\sqrt{\sup_{[u',u_1]}{F}}\right)}=\sqrt{\sup_{[u,u_1]}{F}}.
\end{equation}

 We obtain:

\begin{align}
F(u) \le &2C+ D\sqrt{\sup_{[u,u_1]}{F}}+D \sqrt{\sup_{[u,u_1]}{F}}(u_1-U)B\exp(B(u_1-U)) \nonumber \\
&+2C (u_1-U)B\exp(B(u_1-U)).
\end{align}

Therefore, there exist contants $A'$, $B'$, independent of $u$ and $v$ such that:

\begin{equation}
F(u) \le A'+B'\sqrt{\sup_{[u,u_1]}{F}}.
\end{equation}
This last inequality is true in particular at every $u_m$ where $F(u)$ reaches its maximum in $[u_m,u_1]$:

\begin{equation}
F(u_m) \le A'+B'\sqrt{F(u_m)},
\end{equation}
from which we obtain that $\sqrt{F(u_m)} \le \frac{B'+\sqrt{B'^2+4A'}}{2}$ and by definition of $u_m$, we obtain an upper bound on $F$, which implies an upper bound on G.

 This uniform bound on $G$ gives us uniform bounds on the first derivatives of $U$ and this implies that we can extend $U$ continuously to the past boundary of $\mathcal{F}$ in $\mathcal{M}'$. We can then use the uniform bound on $G$ to get uniform bounds on the first derivatives of $A$ and extends $A$ continuously.

\subsection{Uniform bounds on $\Omega$, $\Omega_u$, $\Omega_v$} 
Observing that the integrals of the twist quantities and the matter terms that appear in (\ref{ee:wo}) can be bounded using (\ref{ee:wr}), we obtain bounds on $\Omega_u/\Omega$, $\Omega_v/\Omega$ and then bounds above and below on $\log \Omega$. In particular we can extend $\Omega$ continuously to the past boundary of $\mathcal{F}$ to a stricly positive function.

\section{The vacuum case} \label{se:vc}
In the following section, we will complete the proof of Theorem \ref{maintheorem}. 
In the vacuum case, the right-hand sides of all the auxiliary equations vanish. This implies that by choosing a linear combination of $X$ and $Y$, we can ensure that $\Gamma=0$ and $\frac{\Pi}{\Omega^2}r^3 e^{-2U}=K$ for some constant $K$. Note that in this case, the terms containing the twist quantities in equations (\ref{ee:wu}) and (\ref{ee:wa}) vanish. Moreover, in the fundamental domain $\mathcal{F}$ of section \ref{se:es}, the terms containing the twist quantities in equation (\ref{ee:wr}) are bounded. We therefore obtain bounds on $U_{uv}$, $A_{uv}$ and $r_{uv}$. Finally, since there are no more matter terms in the right-hand sides of (\ref{ee:ru}) and (\ref{ee:rv}), we obtain bounds on $r_{uu}$ and $r_{vv}$.

We shall first show that $A_v=U_v=0$ on the past boundary before extending the estimates to the tip of the universal cover. The inextendibility criterion will then follow easily.

\subsection{Uniform bounds on $A_{vv}$, $U_{vv}$} \label{es:c2}
To get the bounds on $A_{vv}$, $U_{vv}$, we first take the $v$ derivative of (\ref{ee:wu}) and (\ref{ee:wa}):
\begin{align} \label{der:wave}
A_{uvv}=&-2 A_{vv}U_u+A_uU_{vv}+\frac{r_u}{2r}A_{vv}+\phi(u,v), \\
U_{uvv}=&-\demir \nu U_{vv}+\demirr e^{4U} A_u A_{vv}+ \psi(u,v),
\end{align}
where $\phi$ and $\psi$ contain some previously bounded functions. Choose a null ray $v=v'$ belonging to $\mathcal{F}$. The bounds previously found are valid along this ray. We can consider the system (\ref{der:wave}) as a differential equation in $u$ for the vector $\left( \begin{array}{c} A_{vv} \\ U_{vv} \end{array} \right)$:

\begin{equation}
\frac{d}{du}\left( \begin{array}{c} A_{vv} \\ U_{vv} \end{array} \right)=B \left( \begin{array}{c} A_{vv} \\ U_{vv} \end{array} \right)+\left( \begin{array}{c} \phi \\ \psi \end{array} \right),
\end{equation}
where $B$ is $2 \times 2$ matrix whose coefficients are bounded. We can now integrate the last equation, take the norms and use Gronwall's lemma to get the bounds on $A_{vv}$, $U_{vv}$.

\comment{In fact, it is easy to see that we can repeat the same argument as many times as the degree of differentiablity of A allowed us to do. }

\subsection{Value of $\lambda$, $A$, $U$, $A_v$, $U_v$, $A_{vv}$, $U_{vv}$ on the past boundary}

Since $r$ has been shown to have at least a $C^1$ extension to the boundary and is constant on $u=U$, we have $\lambda=0$ on $u=U$. This implies from the Raychaudhuri equation (\ref{ee:rv}) that $A_v$=$U_v$=0.
Repeating the argument of section \ref{es:c2}, we obtain bounds on the first derivatives of $A_{vv}$ and $U_{vv}$. Since $A_v$ and $U_v$ are constant on $u=U$, we obtain that $A_{vv}$ and $U_{vv}$ extend continuously to $0$ on $u=U$.

\subsection{Periodicity}
Applying the deck transformation to the original fundamental domain, we can bound the same quantities \footnote{The bounds on $A$ and $U$ do not depend on the fundamental domain by periodicity but all other bounds depend on the fundamental domain because of the broken periodicity.} in any fundamental domain and obtain in particular that $A$, $U$ as well as their first and second $v$-derivatives have continuous extension on $u=U$, with $A_v=U_v=A_{vv}=U_{vv}=0$.

\subsection{Extension to the tip of the universal cover} \label{se:tip}
We can now apply a similar method while changing the initial data to obtain estimates for the tip of the universal cover. 

For $p=(u,v) \in \mathcal{T}$, we integrate again equations (\ref{eq:ham}) and (\ref{eq:lag}) along constant $v$ and constant $u$ rays\comment{, with $v \in [v,v_1]$ and $u \in [U,u]$}:
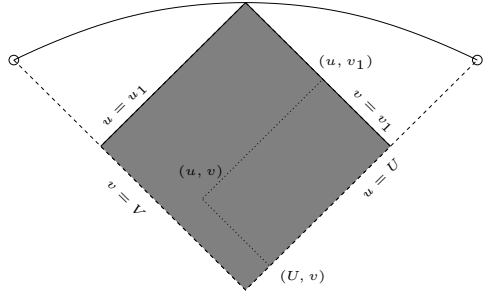
\[ \input{penrose_diagram_integration_tip.pstex_t} \]

We obtain:
\begin{align}
[G+H](u,v)&=[G+H](U,v)-\int^{u}_{U} \left( 2U_uU_v \lambda + \lambda \frac{e^{4U}}{2r^2} A_u A_v \right) \nonumber \\
&+\int^{u}_{U}rU_v \Omega^2 \Lambda,  \\
[G-H](u,v)&=[G-H](u,v_1)+\int^{v_1}_{v}\left(2U_uU_v \nu - \nu \frac{e^{4U}}{2r^2} A_u A_v \right) \nonumber\\
&-\int^{v_1}_{v} rU_u \Omega^2 \Lambda.
\end{align}

Similarly to the estimates of section \ref{es:AU}, we obtain:

\begin{align}
[G+H](u,v)& \le \int^{u}_{U}\left( \lambda \frac{G}{r} +3\frac{r_{uv}}{\sqrt{r}}(G+1) \right),\\
[G+H](u,v)& \le \int^{u}_{U}\left(\frac{\lambda}{r}+3\frac{r_{uv}}{\sqrt{r}}\right)G du'+\frac{3}{\sqrt{r_0}}\lambda(u,v)
\end{align}
and
\begin{align}
[G-H](u,v) \le \int^{v_1}_{v}\left(\nu \frac{G}{r}+6\frac{r_{uv}}{\sqrt{r}}\sqrt{G} \right)dv'+C,
\end{align}
for some positive constant $C$ depending on the bounds of the metric functions along $v_1$ as derived in section \ref{se:es}.

Therefore, we have:

\begin{equation}
G \le D+\demi \int^{u}_{U}(\frac{\lambda}{r}+3\frac{r_{uv}}{\sqrt{r}})G du'+\demi \int^{v_1}_{v}(\nu \frac{G}{r}+6\frac{r_{uv}}{\sqrt{r}}\sqrt{G}) dv'.
\end{equation}

Since we know that for every $v \in [V,v_0]$, $G(.,v)$ is a continuous function on the compact $[U,u_1]$, we can then apply the method of the end of section \ref{es:AU}, we only need to interchange the role of $u$ and $v$ in the argument.

Therefore, we have derived a uniform bound on $G$ in $\mathcal{T}$, which gives us a bound on $U_{i}$, $U$, $A_{i}$ and $A$. We can now continue as in section \ref{se:es} to get the same estimates. In particular, this implies a continuous extension of $A_u$ to $0$ on $v=V$.
From the Raychaudhuri equations, we obtain bounds on $r_{vv}$, $r_{uu}$ valid in $\mathcal{T}$.

As in section \ref{es:c2}, we can obtain higher order estimates of $A$ and $U$ in every fundamental domain. The continuous extension of $A_{uu}$ and $U_{uu}$ and the fact that $A_u$ is constant on $v=V$ implies that  $A_{uu}=0$ and $U_{uu}=0$ everywhere on $v=V$. Once we have this information, we can apply again energy estimates of the type of section \ref{es:c2} replacing $A_i$ by $A_{ii}$ and $U_i$ by $U_{ii}$ in the definition of $H$ and $G$ and taking the initial data on the past boundary. We obtain this way uniform bounds on the second derivatives of $A$ and $U$ which are independent of the fundamental domain.

From equation \ref{ee:wa}, we obtain $A_{uv}=0$ on the past boundary and therefore $A_{u}=0$, $A_{v}=0$ everywhere on the past boundary using a continuity argument.

\subsection{The contradiction}
Equation (\ref{ee:wa}) together with the $C^1$ initial data $A_u=0$ on $u=U$, $A_v=0$ on $v=V$ form a well-posed characteristic initial-value problem with a homogeneous hyperbolic equation for $A$ and trivial initial data. Standard arguments implies that $A$ is constant everywhere in $\tilde{\mathcal{Q}}$ which is a contradiction in view of our initial assumption. Theorem \ref{maintheorem} is thus proved.

\section{The Vlasov case} \label{se:scm}
We will now proceed to the proof of Theorem \ref{th:vla}. We will therefore suppose that the Vlasov field $f$ does not vanish identically in $(\mathcal{M},g)$, as in the statement of \ref{maintheorem}. Our argument is based on an adaptation to our geometry of the proof of Theorem 13.2 of \cite{dr:ss} and on the estimates of section \ref{se:es}.
In the following the indices $\hat{u}, \hat{v},x,y$ will be used to denote the components of tensors in the $(N_{(u)}, N_{(v)},\drond{}{x},\drond{}{y})$ basis and indices (0,1,2,3) will be use to denote the components of tensors in the basis (\ref{fr:ort}).

Let $\mathcal{F'}$ denote a fundamental domain for $\widetilde{Q}$ bounded by two null curves, $v=v_0$, $v=v_1$, $v_1 \ge v_0$ and let $\mathcal{F}=\mathcal{F'} \cup \tau(\mathcal{F'}) \cup \tau^{-1}(\mathcal{F'})$ where $\tau$ generates the deck transformations. $\mathcal{F}$ is bounded by two null curves, $v=v_2$, $v=v_{-1}$:

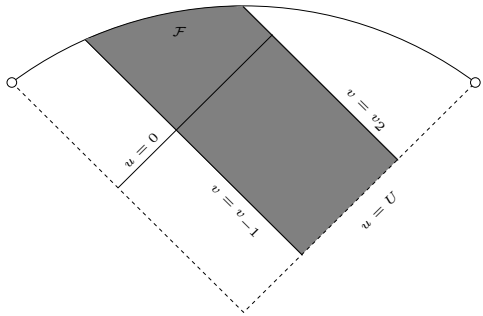
\[ \input{penrose_diagram_with_3_fundamental_domains.pstex_t} \]

By a change of parametrisation, set $\Omega=1$ along $v=v_2$.
Since $T^{\hat{u}\hat{v}}$, $T^{\hat{v}\hat{u}}$, $T^{\hat{u}\hat{u}}$ can be related to the components of curvature in a parallely transported null frame on a null geodesic entering a $C^2$ extension, they can be uniformly bounded along $v=v_2$. We either have $p^{\hat{v}} \ge 1$, in which case, $p^{\hat{v}} \le (p^{\hat{v}})^2$ and therefore  $N^{\hat{v}} \le T^{\hat{v}\hat{v}}$, or $p^{\hat{v}} \le 1$, in which case $p^{\hat{v}} \le (p^{\hat{u}})^2$ and  $N^{\hat{v}} \le T^{\hat{u}\hat{u}}$.
It follows that $N^{\hat{v}}$ is bounded pointwise and thus the flux through $v=v_2$ is bounded. By periodicity, the flux through $v=v_{-1}$ is also bounded. From these bounds and conservation of particle current, it follows that the particle flux is uniformly bounded along any constant $u$ null ray in $\mathcal{F}$ and approaches the initial flux through $\mathcal{F} \cap \Sigma$, as $u \rightarrow U$. Since $f$ is not identically zero in $(\mathcal{M},g)$, we have:

\begin{equation} \label{in:flux}
\lim_{u \rightarrow U} \int^{v_1}_{v_0} N^{\hat{u}}r \Omega^2 dv > \delta_0.
\end{equation}

Since 

\begin{eqnarray}
\lim_{u \rightarrow U}\int^{v_0}_{v_{-1}}\lambda(u,v)dv&=0, \nonumber \\
\lim_{u \rightarrow U}\int^{v_2}_{v_1}\lambda(u,v)dv&=0, \nonumber \\
\end{eqnarray}
there must exist for every $u > U$, $\tilde{v}_0(u)$ and $\tilde{v}_1(u)$ where $v_{-1} \le \tilde{v}_0(u) \le v_0$, $v_1 \le \tilde{v}_1(u) \le v_2$, such that

\begin{equation}
\lim_{u\rightarrow U}\lambda(u,\tilde{v}_1(u))=0.
\end{equation}

Note that $\alpha'(v)$ is bounded in $\mathcal{F}$ since it is a continuous function on the compact $[v_1,v_2]$.
Integrating equation (\ref{ee:rv}) and using the bounds on $\Omega$, $U$ and $r$, $\alpha'$, we obtain that:

\begin{equation} \label{lim:vv}
\lim_{u \rightarrow U}\beta'(u) \int^{v_1}_{v_0}(\rho+P_1-2J_1)dv \le \lim_{u \rightarrow U} \beta'(u)\int^{\tilde{v_1}(u)}_{\tilde{v_0}(u)}(\rho+P_1-2J_1)dv=0.
\end{equation}

We will now use a result obtained in section 3 of \cite{ar:cmc2ds}. If we consider the Vlasov field $f$ as a function of the space-time coordinates and $(p_1, p_2, p_3)$, where the $p_i$ are the components of the momentum of the geodesics in the orthonormal frame (\ref{fr:ort}), the support of the Vlasov $f$ in $p^2$, $p^3$ is bounded as long as $U$ and $A$ are bounded:

\begin{equation}
\sup \left \{ |p^2|, |p^3|/ \exists (u,v,p^1) / f(u,v,p^1,p^2,p^3) \neq 0 \right\} \le M.
\end{equation}

Using this and the fact that $p^0-p^1 \ge \frac{\delta}{\sqrt{\beta'}}$ implies $\sqrt{\beta'}(p^0-p^1) \le \frac{\beta'}{\delta}(p^0-p^1)^2$, we obtain:
\begin{eqnarray}
N^{\hat{u}}&=&\frac{\sqrt{\beta'}}{\sqrt{\alpha'}\Omega}\left( N^0-N^1 \right), \nonumber \\
&=& \frac{\sqrt{\beta'}}{\sqrt{\alpha'}\Omega} \Bigg( \int_{(p^0-p^1) \ge \frac{\delta}{\sqrt{\beta'}}}\frac{p^0-p^1}{p^0}dp^1dp^2dp^3 \nonumber \\
 &&+\int_{(p^0-p^1) \le \frac{\delta}{\beta'}}\frac{p^0-p^1}{p^0}dp^1dp^2dp^3 \Bigg), \nonumber \\
N^{\hat{u}}&\le& \delta^{-1}A \int f\frac{(p^0-p^1)^2}{p^0}dp^1dp^2dp^3+B \delta.
\end{eqnarray}
where $A$ and $B$ are constants which depend on $M$ and the strictly positive lower bounds on $\alpha'(v)$ and $\Omega$.

From (\ref{lim:vv}), the first term on the right-hand side goes to zero as $u \rightarrow U$ and therefore choosing $\delta$ small enough contradicts (\ref{in:flux}).
Theorem \ref{th:vla} has been therefore proved.

\section{Comments}
While we still do not know whether, generically, $r$ goes to $0$ as one approaches the past boundary of $T^2$-symmetric spacetimes with positive cosmological constant and with our without collisionless matter, our result shows that such knowledge is actually unnecessary as far as the $C^2$ formulation of strong cosmic censorship is concerned for the Einstein-Vlasov system. Indeed, Corollary \ref{completevlasov} completes the proof of the $C^2$ formulation of strong cosmic censorship for this class of spacetimes. Moreover, Theorem \ref{maintheorem} means that we only need to focus on the cases where $r$ goes to $0$ for the vacuum models. Note that the set of $T^2$-symmetric spacetimes with positive cosmological constant for which $r_0>0$ is non-empty. We describe briefly how one can build a family of solutions in appendix \ref{ap:1pf}.

It should be emphasized that the positivity of the cosmological constant plays an important role in our analysis. In particular, once we know that a possible Cauchy horizon needs to be regular everywhere and must include at least one side of the past boundary, as shown in section \ref{se:es}, the Einstein equation (\ref{ee:wr}) implies that the integral curve of the null generator of the horizon $u=U$ is past incomplete, or equivalently, that the horizon is non-degenerate. This is an interesting fact since understanding the degeneracy of possible horizons is often thought to be difficult. In the vacuum case, the estimates of section \ref{se:vc} shows that the Cauchy horizon must then coincide with the past boundary of the maximal Cauchy development. We therefore have a bifurcate Cauchy horizon which is similar to the birfucate horizons of \cite{frw:rtsh}.

The techniques used to build the appropriate initial data for the estimates of section \ref{se:es} have relied heavily on the continuity of the curvature tensor. Therefore the methods used here are unlikely to be extended to the $C^0$ formulation of strong cosmic censorship \cite{ch:givp, md:ssems, md:icbh}.

\section{Acknowledgements}
I would like to thank Mihalis Dafermos for suggesting this problem as well as for many useful advice and comments. I would also like to thank Alan Rendall for pointing out the existence of the example of appendix \ref{ap:1pf} and John Stewart for his help along the way. Finally, I wish to gratefully acknowledge funding from EPSRC.

\bibliography{bibliog}

\appendix

\section{The Vlasov equation} \label{ap:ve}
Let $\mathcal{P} \subset \mathcal{TM}$ denote the set of all future directed timelike vectors of length $-1$. $\mathcal{P}$ is classically called the $\emph{mass shell}$. Let $f$ denote a nonnegative function on the mass shell. The Vlasov equation equation for $f$ is derived from the condition that $f$ be preserved along geodesics. In coordinates, we therefore have:

\begin{equation}
p^\alpha \partial_{x^\alpha}f-\Gamma^{\alpha}_{\beta \gamma}p^{\beta}p^{\gamma}\partial_{p^\alpha}f=0,
\end{equation}
where $p^{\alpha}$ denotes the momentum coordinates on the tangent bundle conjugate to $x^\alpha$. Moreover, in the case of $T^2$-symmetry, $f$ is assumed to be invariant under the action induced on $\mathcal{P}$ by the action of $T^2$ on $\mathcal{M}$.

The energy momentum tensor is defined by:

\begin{equation} \label{em:tmn}
T_{\alpha \beta}=\int_{\pi^{-1}(x)}p_\alpha p_\beta f,
\end{equation}
where $\pi : \mathcal{P} \rightarrow \mathcal{M}$ is the natural projection from the mass shell to the spacetime and the integral is with respect to the natural volume form on $\pi^{-1}(x)$.

In an orthonormal frame, with $p^0=\sqrt{1+\delta_{ij}p^i p^j}$, the components of the energy-momentum tensor are given by:
\begin{eqnarray}
\rho(u,v)&=& \int_{\mathbb{R}^3} p^0 f(u,v,p)d^3p, \label{em:rho} \\
P_k(u,v)&=& \int_{\mathbb{R}^3}\frac{(p^k)^2}{p^0}f(u,v,p)d^3p,\\
J_k(u,v)&=& \int_{\mathbb{R}^3}p^k f(u,v,p)d^3p, \\
S_{jk}(u,v)&=&\int_{\mathbb{R}^3}\frac{p^jp^k}{p^0}f(u,v,p)d^3p, \label{em:S}
\end{eqnarray}

Both the \emph{dominant} and the \emph{strong energy conditions} are valid for Vlasov matter as a direct consequence of the definitions.

\section{The class of initial data for the Vlasov field} \label{ap:idv}

We will require that $f$ has initially compact support in $p^2$ and $p^3$. Since $A$, $U$ and $r$ are bounded on the initial Cauchy surface, it is equivalent to say that $f$ has initially compact support in $p^x$ and $p^y$. Note that this requirement is compatible with assumption 1 of Corollary \ref{completevlasov} since we do not add any constraint on the support of $f$ in $p^0$ and $p^1$.

\section{Conservation of particle current}
Define the particle current vector field $N$ by:

\begin{equation}
N^{\alpha}= \int_{\pi^{-1}(x)}p_\alpha f.
\end{equation}

The Vlasov equation implies that $N$ is divergence free and we obtain the conservation law:

\begin{eqnarray}
\int^{u_2}_{u_1}N^{\hat{v}}\Omega^2r^2(u,v_1)du&+&\int^{v_2}_{v_1}N^{\hat{u}}\Omega^2r^2(u_1,v)dv=\int^{u_2}_{u_1}N^{\hat{v}}\Omega^2r^2(u,v_2)du \nonumber \\&+&\int^{v_2}_{v_1}N^{\hat{u}}\Omega^2r^2(u_2,v)dv,
\end{eqnarray}
where $\hat{v}$, $\hat{u}$ are indices for the components of tensors along the null vectors $N_{(u)}$, $N_{(v)}$ in the basis $(N_{(u)}, N_{(v)},\drond{}{x}, \drond{}{y})$.

\section{Change of global null coordinates on $\mathcal{\widetilde{Q}}$ } \label{ap:gnc}
In this appendix, we recall how rescaling of global null coordinates affect the metric functions and the energy-momentum components. Suppose therefore that $(u,v) \in (U,u_0] \times (V,v_0]$ is a global null coordinate system on $\mathcal{\widetilde{Q}}$, where $u_0$ and $v_0$ are real numbers and $U$ and $V$ are real numbers or $-\infty$.

 Let $f$ and $g$ be smooth functions on $(U,u_0] \times (V,v_0]$ with $f' > 0$ and $g'>0$. Let $(u^*(u),v^*(v))=(f(u),g(v))$. In the coordinate system defined by $(u^*,v^*)$, the functions $\Omega^*$, $U^*$, $A^*$, $r^*$ are given by:

\begin{eqnarray}
{\Omega^*}^2&=&\frac{\Omega^2}{f'g'}, \\
U^*&=&U, \\
A^*&=&A, \\
r^*&=&r. 
\end{eqnarray}
With $\Gamma$ and $\Pi$ defined by (\ref{eq:gamma}) and (\ref{eq:pi}), we have that $\frac{\Gamma^*}{{\Omega^*}^2}=\frac{\Gamma}{\Omega^2}$ and  $\frac{\Pi^*}{{\Omega^*}^2}=\frac{\Pi}{\Omega^2}$.
The derivatives change in the obvious way, for instance $r^*_{u^*}=\frac{r_u}{f'}$, and since $f$ and $g$ are functions of a single variable, $r^*_{u^*v^*}=\frac{r_{uv}}{f'g'}$. 
Finally, since $\rho$, $P_k$, $S_{ik}$ are defined in the fixed frame (\ref{fr:ort}), they are left unchanged by the change of coordinates.

\section{A family of solutions with $r_0 > 0$} \label{ap:1pf}
We consider vacuum $T^2$-symmetric spacetimes with positive cosmological constant and assume that the metric functions  independent of $\theta$. Moreover, we consider the ansatz $r=e^{2U}$, $G=H=A=0$.
 Working in areal coordinates ($t=r$), the metric can be written as:

\begin{equation}
ds^2=-\frac{e^{2 \gamma}}{t}\delta dt^2+\frac{e^{2 \gamma}}{t} d \theta^2+t(dx^2+dy^2).
\end{equation}
The Einstein equations reduce to the system:
\begin{eqnarray}
\gamma_t&=&\frac{1}{4t}+\delta \Lambda e^{2\gamma}, \nonumber \\
\delta_t&=&-4 \delta^2 e^{2 \gamma} \Lambda. \label{eq:delta}
\end{eqnarray}

From which we obtain:

\begin{eqnarray}
\frac{\delta_t}{\delta}&=&-4(\gamma_t-\frac{1}{4t}), \\
\frac{\delta_t}{\delta}&=&-4 \gamma_t +\frac{1}{t}, \\
\ln \delta&=&-4\gamma+\ln t +2 \ln A,
\end{eqnarray}
where $A$ is a stricly positive constant.

\begin{eqnarray}
4 \gamma &=& \ln t - \ln \delta+2 \ln A, \\ 
2 \gamma &=& \frac{1}{2} \ln t-\frac{1}{2} \ln \delta+\ln A, \\
e^{2 \gamma} &=& A \sqrt{\frac{t}{\delta}}. \label{eq:link}
\end{eqnarray}

Inserting (\ref{eq:link}) in (\ref{eq:delta}), we obtain:

\begin{eqnarray}
\delta_t &=& - 4 \delta^2 A \sqrt{\frac{t}{\delta}} \Lambda, \\
\delta_t &=& -4 \delta^{3/2} A \Lambda \sqrt{t}, \\
-\demi \frac{\delta_t}{\delta^{3/2}}&=& 2 A \Lambda \sqrt{t}
\end{eqnarray}
and by integration:
\begin{eqnarray}
\frac{1}{\sqrt{\delta}}&=& \frac{4}{3} A \Lambda t^{3/2}+C,
\end{eqnarray}
where the constant $C$ satisfies:

\begin{eqnarray}
\frac{4}{3} A \Lambda t^{3/2}+C &>& 0, \\
C &>& -\frac{4}{3} A \Lambda t^{3/2}.
\end{eqnarray}

We obtain for the solution:

\begin{eqnarray}
\delta &=&\frac{1}{\left( \frac{4}{3} A \Lambda t^{3/2}+C \right)^2}, \\
e^{2 \gamma} &=& A \sqrt{\frac{t}{\delta}} = A\left(\frac{4}{3} A \Lambda t^2+C \sqrt{t}\right),
\end{eqnarray}

or for the metric components:

\begin{eqnarray}
\frac{e^{2 \gamma}}{t}&=&A\left(\frac{4}{3} A \Lambda t+\frac{C}{\sqrt{t}}\right), \\
\frac{e^{2 \gamma}}{t}\delta &=&A \frac{1}{\frac{4}{3} A \Lambda t^2+C \sqrt{t}}.
\end{eqnarray}

Suppose $C < 0$ and let $D=-C$.
The previous coordinate system breaks down at $t_0$:

\begin{equation}
t_0= \left(\frac{3}{4}\frac{D}{A \Lambda}\right)^{2/3}.
\end{equation}

However, this system of coordinates is known to cover the whole past maximal Cauchy development of the initial data. Therefore, the area element of such solutions does not go to zero on the past boundary of the maximal Cauchy development.
Nevertheless, one should note that these solutions have a Cauchy horizon at $t=t_0$, as one can easily get around the coordinate singularity with a coordinate transformation of the type $t= f(t)+\theta$, with $f'(t)=\pm \frac{1} {A \left( 4/3 A \Lambda t+ \frac{C}{\sqrt{t}} \right)}$. These solutions are clearly excluded by the generic condition of Theorem \ref{maintheorem}.

\end{document}